\documentclass{moriond}
\usepackage{subcaption}
\usepackage{graphicx}
\usepackage{amssymb,amsmath,verbatim,mathtools,needspace,enumitem,etoolbox,graphicx,physics,microtype,afterpage,bigints,gensymb,tabularx,mathtools}
\usepackage[dvipsnames, usenames]{xcolor}
\definecolor{linkcolor}{rgb}{0.0,0.3,0.5}
\definecolor{dodgerblue}{HTML}{1E90FF}
\usepackage{scalerel}
\usepackage{orcidlink}
\usepackage{stackengine,wasysym}

\makeatletter
\newcommand*{\balancecolsandclearpage}{\close@column@grid \cleardoublepage \twocolumngrid}
\makeatother
\usepackage{lipsum}

\def\be{\begin{equation}}
\def\ee{\end{equation}}
\def\bea{\begin{eqnarray}}
\def\eea{\end{eqnarray}}

\newcommand{\Photo}{}

\begin{document}
\vspace*{4cm}
\title{Up-down binaries are unstable and we want to know}

\author{Viola De Renzis$\,$\orcidlink{0000-0002-0933-3579}$\,^{1,2}$}

\address{ \vspace{0.2cm}
 \footnotesize
$^1$ Dipartimento di Fisica ``G. Occhialini'', Universit\'a degli Studi di\\Milano-Bicocca, Piazza della Scienza 3, 20126 Milano, Italy \\
$^2$ INFN, Sezione di Milano-Bicocca, Piazza della Scienza 3, 20126 Milano, Italy}

\maketitle\abstracts{
The relativistic spin-precession equations for black-hole binaries have four different equilibrium solutions that correspond to systems where the two individual black hole spins are either aligned or anti-aligned with the orbital angular momentum. 
Surprisingly, it was demonstrated that only three of these equilibrium solutions are stable. Binary systems in the up--down configuration, where the spin of the heavier (lighter) black hole is co- (counter-) aligned with the orbital angular momentum, might be unstable to small perturbations of the spin directions.  After the onset of the up--down instability, that occurs after a specific critical orbital separation $r_\mathrm{UD+}$, the binary becomes unstable to spin precession leading to large misalignment of the spins. In this work, we present a Bayesian procedure based on the Savage-Dickey density ratio to test the up–down origin of gravitational-wave events. We apply this procedure to look for promising candidates among the events detected so far during the first three observing runs performed by LIGO/Virgo. 
}

\section{Introduction}
One of the main goal of gravitational-wave (GW) observations is to measure the masses and the spins of astrophysical black holes (BHs). 
The study of the spin dynamics is a fundamental ingredients in the population framework since spin orientations constitute a signature to discriminate between the two main formation channels of stellar-mass black-hole binaries.~\cite{arXiv:1302.4442,arXiv:1609.05916,arXiv:1503.04307,arXiv:1703.06873}  
These systems can originate through either the isolated evolution of massive binary stars or dynamical encounters in dense stellar clusters~\cite{MF}. In the first case,  the two BH spins $\boldsymbol{S_{1,2}}$ are expected to be preferentially aligned with the orbital angular momentum $\boldsymbol{L}$ of the binary. On the other hand, the spin vectors of dynamically formed binary BHs are predicted to show an isotropic distribution that might lead to detectable spin-precession modulations in the emitted GWs.

The four configurations in which the BH spins are either aligned or anti-aligned with  $\boldsymbol{L}$ are equilibrium solutions to the relativistic spin-precession equations. We conventionally call them \textit{up–up}, \textit{down–down}, \textit{down–up}, and \textit{up–down} configurations, where ``up'' (``down'') indicates BH spins that are aligned (anti-aligned) with $\boldsymbol{L}$ and the direction before (after) the hyphen refers to the more (less) massive BH. 
It was shown~\cite{arXiv:1506.09116} that if a small perturbation in the spin direction is applied, only \textit{up–up}, \textit{down–down}, \textit{down–up} binaries remain stable while BHs in the \textit{up–down} configuration might become unstable to spin precession. These sources encounter an instability at the critical orbital separation

\begin{equation}
r_\mathrm{UD+} = \frac{\left(\sqrt{\chi_1}+\sqrt{q\chi_2}\right)^4}{(1-q)^2}M\,,
\label{eq:rplus}
\end{equation}
where $\chi_{i}=S_{i}/m_{i}^2$ are the Kerr parameters of the BHs, $q=m_2/m_1\leq 1$ is the mass ratio, and $M=m_1+m_2$ is the total mass of the system.
Surprisingly, it was discovered~\cite{arXiv:2003.02281}, that,,
rather than dispersing in a random point of the parameter space, unstable up--down binaries tend to reach a specific region
at the end of their evolution. The endpoint of the up--down instability is a precessing configuration defined by the following three conditions
\begin{align}
\cos\theta_1&=\frac{\chi_1-q\chi_2}{\chi_1+q\chi_2}\,, \label{eq:tilt1} \\
\cos\theta_2&=\frac{\chi_1-q\chi_2}{\chi_1+q\chi_2}\,, \label{eq:tilt2}\\
\phi_{12}&=0\,,\label{eq:phi12}
\end{align}
where $\theta_{i}$ are the tilts angles between $\boldsymbol{S}_i$ and $\boldsymbol{L}$, and  $\phi_{12}$ is the azimuthal angle between the two BH spins measured in the orbital plane. From now on, we define $\cos\theta_\mathrm{UD} = ({\chi_1-q\chi_2})/({\chi_1+q\chi_2})$.

\section{Methods}
We apply a model-selection analysis using the Savage-Dickey ratio approximation and calculate the posterior odds between a broader model $\mathcal{H}_\mathrm{B}$ describing  the case of generically precessing binaries and a narrower model $\mathcal{H}_\mathrm{N}$ representing the up--dow hypothesis. 
We assume equal model priors such that the posterior odds reduce 
to the Bayes factor $\mathcal{B}$. According to the Jeffrey scale~\cite{Jeffrey}, $ | \ln \mathcal{B}|<1$ is classified as ``inconclusive,'' $1< |\ln \mathcal{B}| <2.5$ is classified as as ``weak'' evidence, $2.5<|\ln \mathcal{B}|<5$ is classified as ``moderate'' evidence, and $|\ln \mathcal{B}|>5$ is classified as ``strong'' evidence.

In particular, the narrow model $\mathcal{H}_\mathrm{N}$ is nested in $\mathcal{H}_\mathrm{B}$, meaning that among the whole set of 15 parameters $\theta=\{\varphi,\gamma\}$ describing a binary BH merger, a subset of 12 parameters $\varphi$ is common to both models (including BH masses, dimensionless spin magnitudes , sky location ...). The other parameters $\gamma=\{\cos\theta_1,\cos\theta_2,\phi_{12}\}$ are allowed to vary in the wider model but are fixed to the three constraints defined in Eqs.~(\ref{eq:tilt1}--\ref{eq:phi12}) in the narrow model. Let us also assume that the prior on $\varphi$ is the same for the two models. Under these assumptions and adopting~\cite{arXiv:2304.13063v1} a convenient reparametrization $\tilde\gamma$ of Eqs.~(\ref{eq:tilt1}--\ref{eq:phi12}) such that the endpoint of the up-down instability is mapped into $\tilde\gamma=\{0,0,0\}$,
the Bayes factor can be computed through the so-called Savage-Dickey density ratio
\begin{equation}
\mathcal{B}=  \frac{p(\tilde\gamma=0 | d, \mathcal{H}_{\rm B}) }{  \pi(\tilde\gamma=0 | \mathcal{H}_{\rm B})} \;  \,,
\label{eq:BF_SDR}
\end{equation}
where the numerator (denominator) corresponds to the posterior (prior) marginalized over the common parameters $\varphi$. The main advantage of this procedure is that obtaining the Bayes factor only requires evaluating the prior and posterior distributions of the wider model at the point $\tilde\gamma=0$.

\section{Results}
We apply our model-selection analysis to the GW events reported in GWTC-3.~\cite{GWTC1,GWTC2,GWTC3} We analyze a total number of 69 binary BH mergers listed in Table I of Ref~\cite{arXiv:2304.13063v1}.   We select events with false alarm rates $< 1~\mathrm{yr}^{-1}$ in at least one of the LIGO/Virgo searches, excluding those that can potentially include a neutron star. We use data posterior samples using publicly available posterior samples for the GWTC-2.1 and the GWTC-3 data releases. We use data product referring to the {\sc IMRPhenomXPHM} waveform model where the merger rate is uniform in comoving volume and source-frame time.

\begin{figure}
\centering
    \includegraphics[width=0.55\columnwidth]{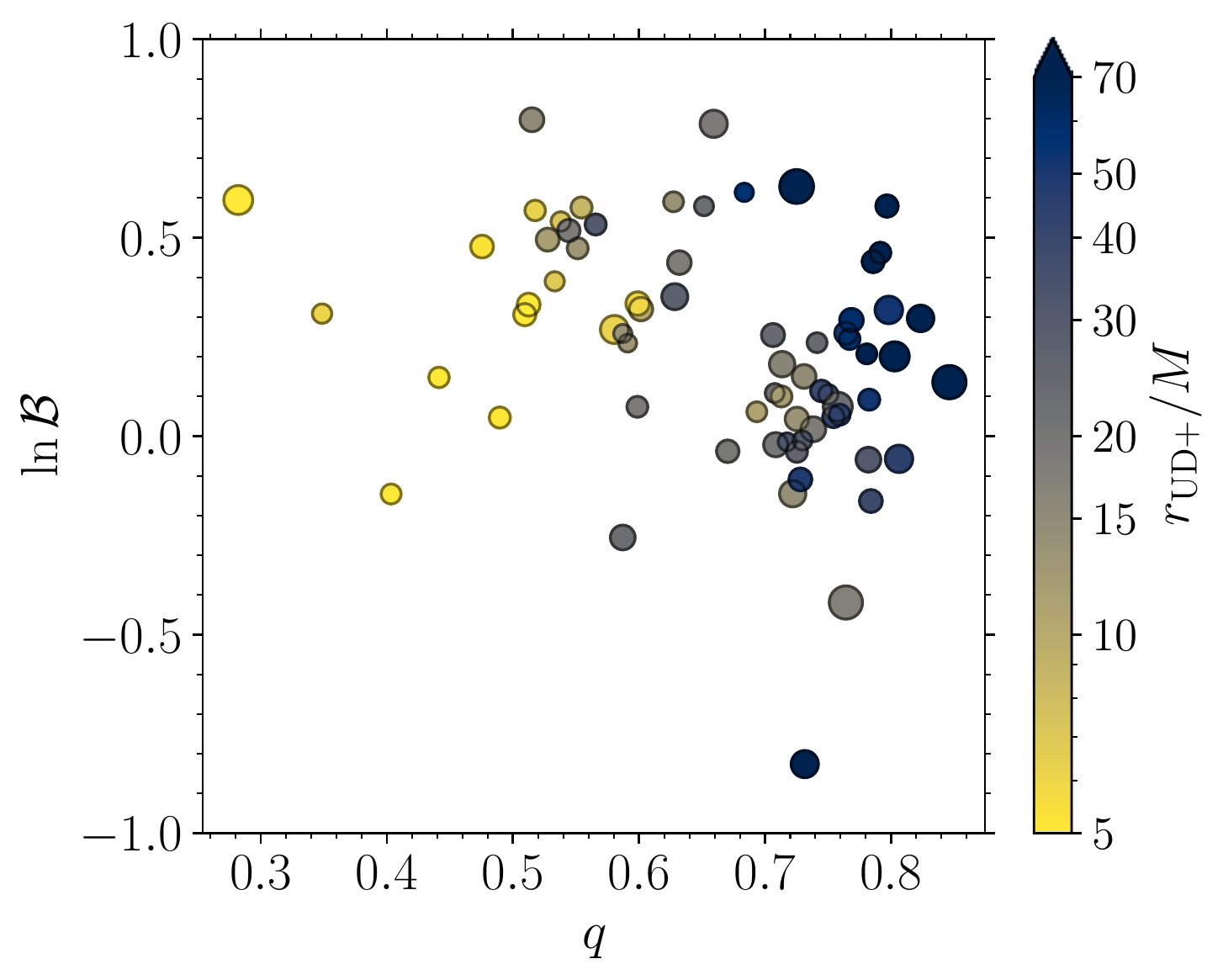}
    \caption{Natural logarithm of the Bayes factor $\mathcal{B}$ as a function of the mass ratio $q$ for 69 GW events selected from GWTC-3. The critical orbital separation $r_\mathrm{UD+}$ is reported on the color bar and the size of each scatter point is directly proportional to the three-detector SNR. 
}
\label{fig:BF_SNR_realdata}
\end{figure}

Our results are shown in Fig.~\ref{fig:BF_SNR_realdata}, where we report the Bayes factor as a function of the mass ratio $q$, the critical separation $r_{\rm UD+}$, and the signal-to-noise ratio (SNR) that is estimated using the median of the optimal network SNR posterior samples. The natural logarithm of the Bayes factor in favor of the up--down model for current GW data is included within the range $\ln\mathcal{B}\in[-0.8,0.8]$, which implies an inconclusive model-selection analysis. This result is expected given the low values of the SNRs with which current GW events have been detected.~\cite{arXiv:2304.13063v1} The current sensitivity of GW detectors do does not allow us to put accurate constraints on the spin magnitudes and orientations, which are key
ingredients to provide discriminative answers in the context of the up--down model selection. The colors of the scatter points in Fig.~\ref{fig:BF_SNR_realdata} show a correlation between $r_{\rm UD+}$ and $q$ which reflects the fact that $r_{\rm UD+}\propto (1-q)^{-2} $; cf. Eq.~(\ref{eq:rplus}).

\begin{figure*}
\centering
\begin{subfigure}{0.45\columnwidth}
    \includegraphics[width=\textwidth]{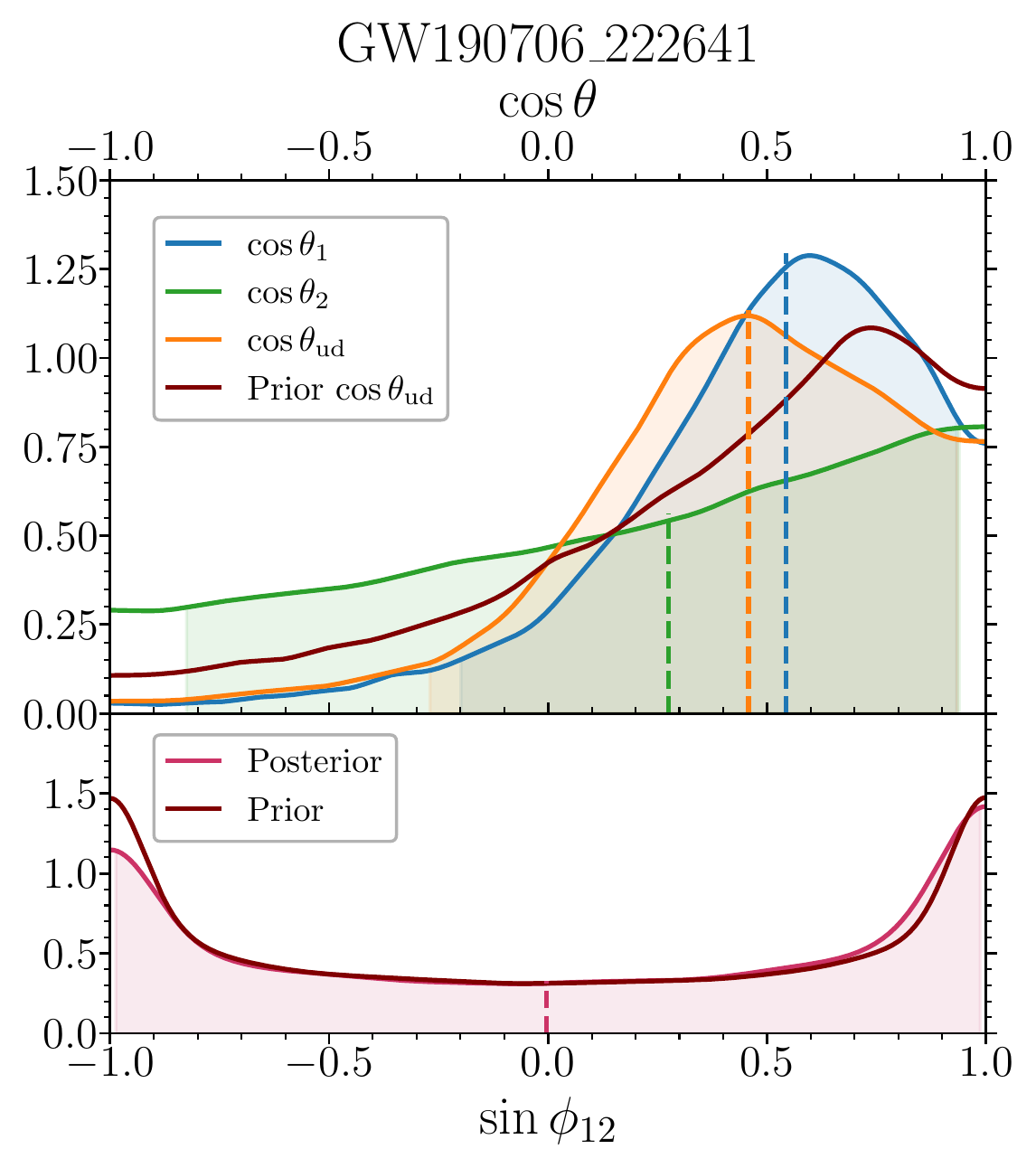}
    \caption{}
    \label{fig:kde}
\end{subfigure}
\hfill
\begin{subfigure}{0.45\columnwidth}
    \includegraphics[width=\textwidth]{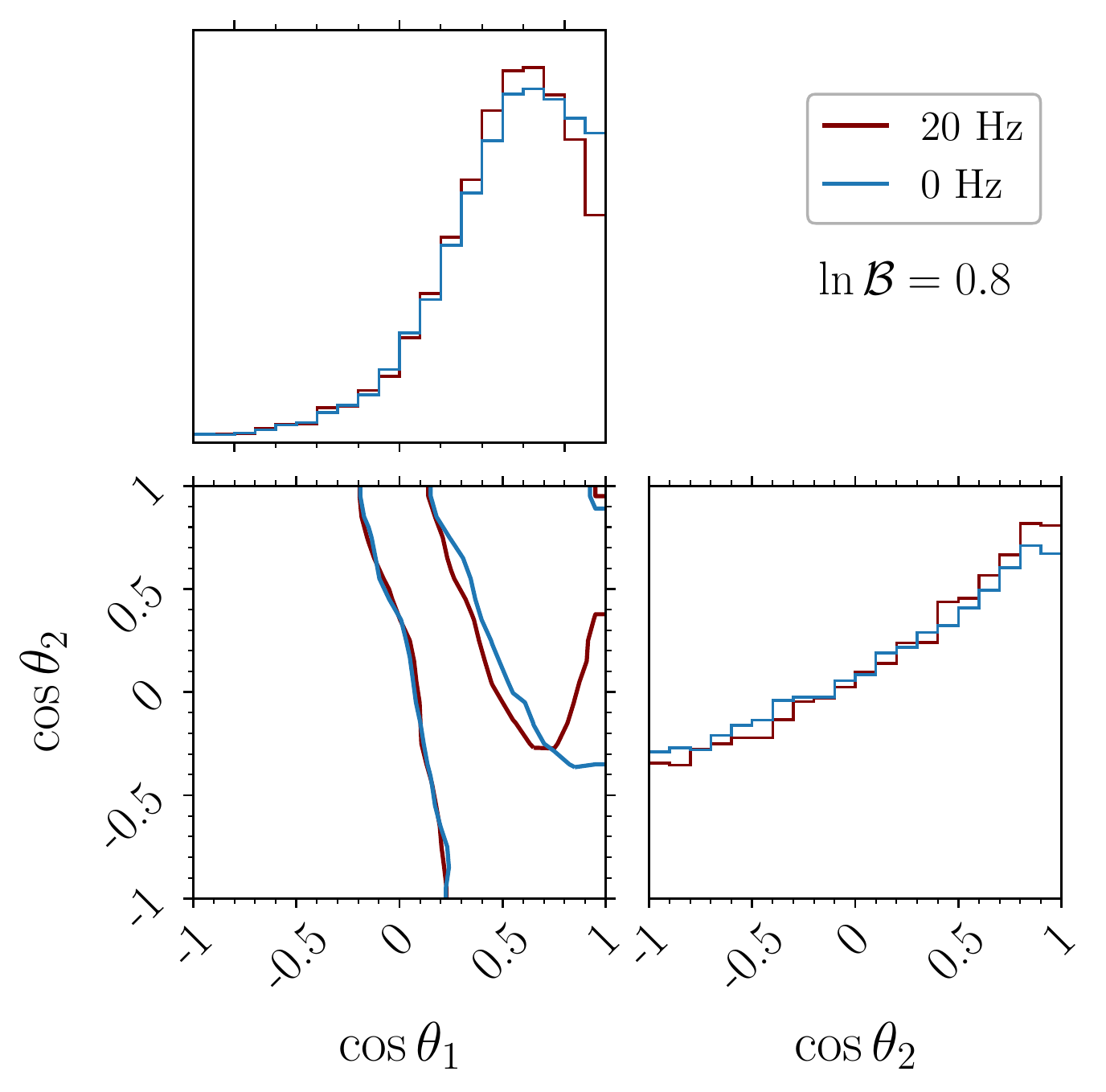}
    \caption{}
    \label{fig:second}
\end{subfigure}     
\caption{\textbf{a)}: The upper subpanel shows the posterior distributions of $\cos\theta_1$ (blue), $\cos\theta_2$(green), $\cos\theta_\mathrm{UD}$ (orange) for the GW event GW190706\_222641. The dark red distribution is the prior of $\cos\theta_\mathrm{UD}$ (dark red), while the prior distributions of $\cos\theta_{1,2}$ are flat. The lower subpanel shows posterior (pink) and prior (dark red) distributions of $\sin\phi_{12}$ for the same event. Dashed vertical lines mark the medians of each distribution while shaded areas indicate the 90\% credible intervals.\\
\textbf{b)}: Joint posterior distribution of the cosine of the tilt angles $\theta_1$ and $\theta_2$ for the GW event GW190706\_222641. Posterior samples are evolved numerically from $f_\mathrm{ref}$= 20 Hz (red) to 0 Hz (blue). Contour levels mark the 50\% and 90\% credible regions.}
\label{fig:figures}
\end{figure*}

Among the set of 69 GW events considered here, we selected the one with the highest value of the Bayes factor, namely GW190706\_222641 for which $\ln\mathcal{B}=0.8$. In the upper
 panel of Fig.~\ref{fig:kde}, we compare the posterior distribution of $\cos\theta_{1,2}$ obtained from the parameter estimation analysis against the theoretical prediction for the up-down instability $\cos\theta_\mathrm{UD}$, which is computed through the posteriors of $q$, $\chi_{1}$ and $\chi_{2}$. The proximity of the median values of these three posterior distributions might be qualitatively interpreted as an indication that the up-down hypothesis is a good description of the observed data. We stress that the appropriate statistical estimator is the Bayes factor, which is in this case is inconclusive.
For completeness, in the lower panel of Fig.~\ref{fig:kde} we show the posterior distribution of $\sin\phi_{12}$, which, according to Eq.~(\ref{eq:phi12}), is predicted to be close to zero 0 for the up–down hypothesis.

The signature imprinted by the up-down instability can be further explored by evolving the posterior distributions of $\cos\theta_{1,2}$ backward in time to infinitely large separations. 
This back-propagation exercise should
illustrate that the posterior samples of  $\cos\theta_{1,2}$ move from the precessing configuration at 20 Hz to the up-down configuration at 0 Hz where  $\cos\theta_{1}=-\cos\theta_2=1$.
The result of the back-propagation for GW190706\_222641 is shown in Fig.~\ref{fig:second}. As expected, given the low values of both the SNR and the Bayes factor, the posterior distributions of $\cos\theta_{1,2}$ at 0 Hz and 20 Hz are largely indistinguishable.

\section{Conclusions}
The main results of our analysis have been presented elsewhere.~\cite{arXiv:2304.13063v1} In this conference contribution, we presented a model selection analysis based on the Savage-Dickey density ratio for the calculation of the Bayes factor and applied it to current GW events in order to identify unstable up-down binary systems. Among the events contained in GWTC-3, we did not find any promising candidate that might be interpreted as a binary systems that was originally stable and aligned in the up– down configuration. 

\section*{Acknowledgments}
V.D.R is supported by European Union's H2020 ERC Starting Grant No.~945155--GWmining, Cariplo Foundation Grant No.~2021-0555, and the ICSC National Research Centre funded by NextGenerationEU. Computational work was performed at CINECA with allocations through INFN, Bicocca, and ISCRA project HP10BEQ9JB.

\end{document}